\documentclass[letterpaper, 10 pt, conference]{ieeeconf}  
\usepackage{graphicx}
\usepackage{xcolor}
\usepackage{listings} 
\usepackage{subfigure}
\usepackage{hyperref}  
\usepackage{stfloats}  
\usepackage{booktabs}
\usepackage{cite}
\let\labelindent\relax
\usepackage{enumitem} 

\usepackage{threeparttable}

\IEEEoverridecommandlockouts                              

\title{\LARGE \bf
SemanticScanpath: Combining Gaze and Speech for Situated Human-Robot Interaction Using LLMs
}

\author{{Elisabeth Menendez$^{1}$, Michael Gienger$^{2}$, Santiago Martínez$^{1}$, Carlos Balaguer$^{1}$, and Anna Belardinelli$^{2}$}
\thanks{$^{1}$Robotics Lab, Department of Systems Engineering and Automation, Universidad Carlos III de Madrid (UC3M),
       {\tt\small \{emenende, scasa, balaguer\}@ing.uc3m.es.}}%
\thanks{$^{2}$Honda Research Institute Europe, Germany, 
    {\tt\small \{firstname.lastname\}@honda-ri.de}}%
\thanks{Supplementary Material:\url{https://hri-eu.github.io/SemanticScanpath/}}%
}

\begin{document}

\maketitle
\thispagestyle{empty}
\pagestyle{empty}

\begin{abstract}
Large Language Models (LLMs) have substantially improved the conversational capabilities of social robots. Nevertheless, for an intuitive and fluent human-robot interaction, robots should be able to ground the conversation by relating ambiguous or underspecified spoken utterances to the current physical situation and to the intents expressed nonverbally by the user, such as through referential gaze. Here, we propose a representation that integrates speech and gaze to enable LLMs to achieve higher situated awareness and correctly resolve ambiguous requests. Our approach relies on a text-based semantic translation of the scanpath produced by the user, along with the verbal requests. It demonstrates LLMs' capabilities to reason about gaze behavior, robustly ignoring spurious glances or irrelevant objects. We validate the system across multiple tasks and two scenarios, showing its superior generality and accuracy compared to control conditions. We demonstrate an implementation on a robotic platform, closing the loop from request interpretation to execution.

\end{abstract}

\section{INTRODUCTION}
Collaborative and assistive robots have been increasingly deployed in public spaces and in contact with non-expert users \cite{holland2021service}.
    Such a target audience requires 
    human-robot communication to occur in the most natural and intuitive way. For this reason, natural language has long been considered a critical modality of interaction with robots \cite{nikolaidis2018planning,MARGE2022101255}, one that 
    comes with
    challenges \cite{reimann2024survey}. While LLMs have opened up a wealth of possibilities for robots to integrate rich conversational capabilities with reasoning and decision-making \cite{li2025embodied,joublin2024copal}, a robot, unlike a virtual assistant (VA) or a conversational agent, is an embodied and situated agent that shares a space with its human partners. In face-to-face interactions between humans, verbal utterances are hardly the only way to communicate \cite{holler2019multimodal}: we assume that our conversation partners share the same perception and understanding of the situation, and they can combine this context with the spoken exchange and related non-verbal cues, such as gestures and gaze. Similarly, a robot partner should be able to ground spoken requests both in the physical situation \cite{tellex2020robots} and in relation to the additional signals sent by the user. LLMs coupled with some form of open-world perception and scene understanding (e.g., \cite{liu2024vision}) can give robots, to some extent, situational awareness by allowing them to associate mentioned nouns with appropriate objects present in the scene. Still, ambiguities might arise when multiple instances of the same object are present or when unspecific references are used (e.g., "Can you put this there?"). Humans make trade-offs between accuracy and efficiency of communication, but in most cases, ambiguities are easily resolved by attending to the speaker's body pose. Gaze, in particular, has long been demonstrated to be a powerful cue to disambiguate referential expressions \cite{HANNA2007596,staudte2008utility} and has been extensively used in human-robot interaction, for establishing joint attention, to direct attention in a deictic way, or to infer intention \cite{admoni2017social,belardinelli2024gaze,menendez2024integrating}.
    \begin{figure}[!t]
      \centering
      \includegraphics[width=0.42\textwidth]{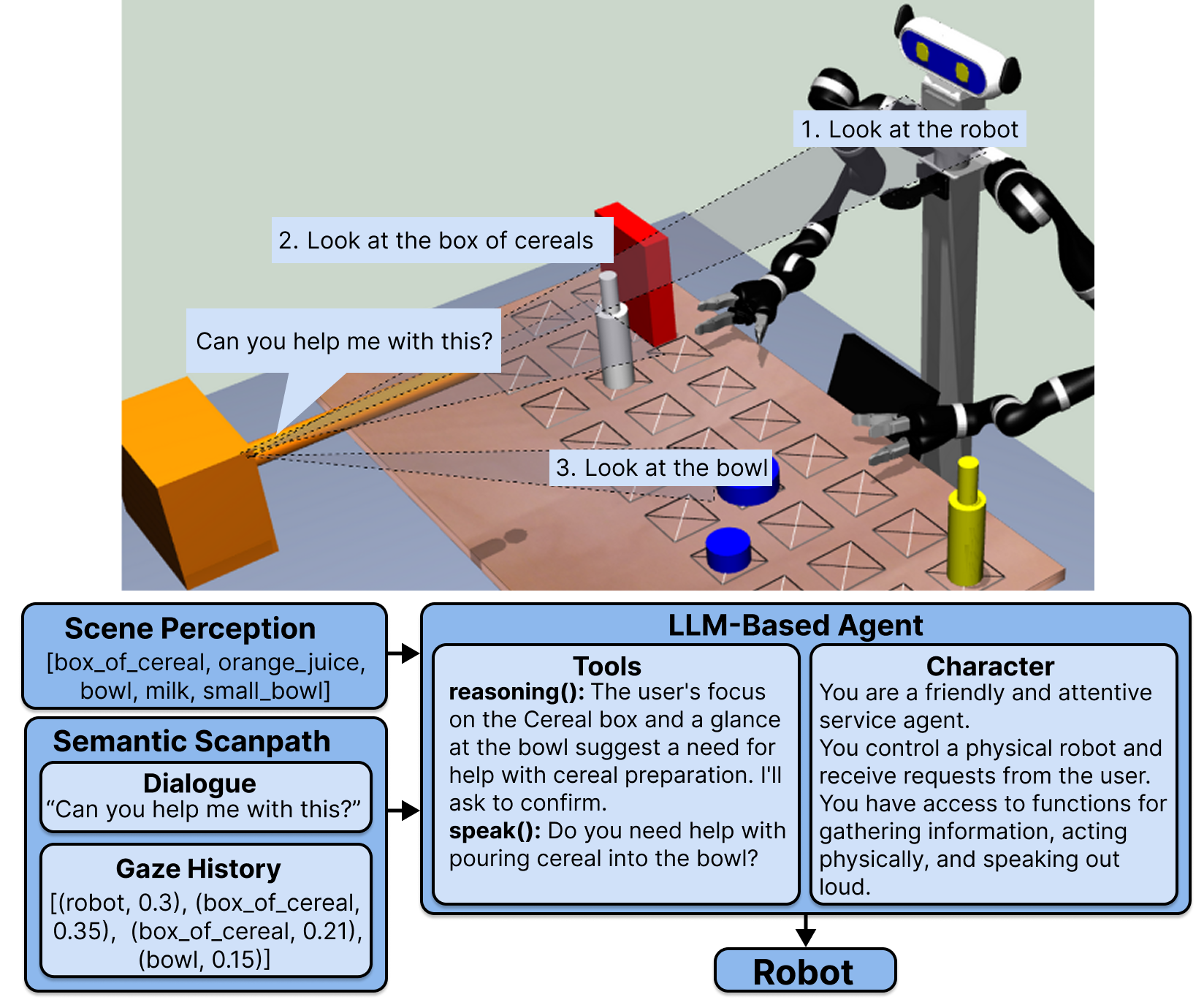}
      \caption{Top: Example interaction, where the user looks sequentially at the robot, the cereal box, and the bowl while speaking. Bottom: Our system architecture, leveraging Semantic Scanpath, a representation combining speech and gaze history to help the LLM-based agent resolve underspecified user requests by reasoning about speech, scene context, and gaze behavior.}
      
      \label{overview}
\end{figure}
    Still, the potential of gaze in complementing verbal input to LLM-based robotic systems has not been much explored thus far. Nevertheless, situated interaction with robots would be greatly improved if it could leverage both enhanced NLP capabilities and appropriate interpretations of non-verbal cues, such as gaze, especially for resolving open-world references \cite{williams2016situated}.

To address this issue, 1) we introduce the Semantic Scanpath, a novel representation that combines a textual representation of the user's gaze path over semantic Areas-of-Interest with transcription of the concurrent utterance; 2) we show that this representation allows LLM-based multi-modal dialogue interaction by relating underspecified user requests to the correct objects while dismissing irrelevant objects 
based on scene context and gaze behavior. Building on prior work on LLM-enhanced robot assistant systems \cite{Tanneberg2024,wang2024lami}, we evaluate the performance with an LLM-enhanced robot assistant 
in different scenarios and across multiple tasks and considering multiple control conditions; 3) we show successful integration on a different robotic platform and discuss the implications of our results and future perspectives for 
natural human-robot interaction.


\section{RELATED WORK}
Human-robot situated interaction can hugely benefit from natural language dialogue. A  critical assumption for natural dialogues in physical contexts is the capacity to infer intentions, even from indirect requests, and to resolve unspecified references \cite{kruijff2010situated}. Before the advent of LLMs, this implied designing probabilistic inference mechanisms and leveraging hierarchical and graph structures to handle references in task-based dialogues \cite{williams2016situated,Chai2014,dogan2024semantically}.
To resolve demonstrative nouns in human-robot collaboration, \cite{wan2022handmethat}
has proposed a benchmark for reinforcement learning deep architectures that can track completed subgoals to predict the current step and disambiguate the requested object. Such dialogue-based approaches typically recognize the complementary role played by multimodal cues \cite{gross2017reliability}, but few have integrated these effectively. LLMs have enabled robotic and virtual assistants to resolve many linguistic challenges \cite{browning2023language}, but situated references in physical interaction still need to rely on scene understanding and non-verbal cues from the speaker. 
In Human-Computer Interaction, contingent-gaze detection has been recently used in combination with LLMs in a deictic way, to inform  VAs of what the user is looking at while asking for information \cite{yan2025voila,lee2024gazepointar,konrad2024gazegpt}
. Similarly, \cite{chang2023specifying} combined gaze pointing and spoken utterances to specify the target of robotic teleoperation. Typically, only the instantaneous point of regard is considered, not the overall gaze behavior (scanpath) during a dialogue turn, which, in its dynamics, can provide a richer picture of the user's intentions and expectations. This was one of the motivations to include such temporal aspects in our work.
It has been further shown that LLM-powered robots are expected to complement their sophisticated language capabilities by also displaying appropriate non-verbal cues \cite{kim2024understanding}. Here, we postulate that robots should also understand non-verbal cues and propose a representation to correctly interpret human gaze behavior in relation to spoken interaction and scene context.


\section{GAZE AND SPEECH DISAMBIGUATION IN SITUATED SCENARIOS}
\label{sec:gaze}
This section formalizes the multimodal grounding problem and presents the proposed architecture for resolving ambiguous spoken requests using gaze and speech (see Fig.~\ref{overview}). The system constructs a textual representation of gaze during the utterance and employs an embodied LLM-based agent to interpret the multimodal input and control the robot.


\subsection{Problem Formulation}
Consider a workspace shared between a robot and a human and containing a set of objects $\mathcal{O}$. During an interaction, the user produces an utterance ${u}$ and exhibits gaze behavior ${g(t)}$, with $t \in W$, where $W$ denotes the temporal window during which the user makes the request. The utterance specifies a request to be carried out in the scene. While the request may describe or imply the action to be performed, the object(s) $\mathcal{O}^* \subseteq \mathcal{O}$ to which it applies may not be uniquely determined from speech alone. The problem addressed in this work is to infer, from $u$, $g(t)$, and $\mathcal{O}$, the object(s) ${\mathcal{O}^*}$ involved in the request, i.e, $\mathcal{O}^* = \Phi(u, g(t), \mathcal{O})$, where ${\Phi(\cdot)}$ denotes the multimodal grounding mechanism.

\subsection{The Semantic Scanpath}

Speech production and gaze behavior are consistently related in a situated interaction. We propose here that they should be jointly evaluated in a suitable representation.  During turn-taking in collaborative scenarios, scanpaths, intended as ordered sequences of fixations, can reveal the start and end of a turn \cite{ho2015speaking}, anticipate objects the user is about to mention \cite{HANNA2007596} and/or manipulate \cite{belardinelli2024gaze}, ultimately making communication less ambiguous and more efficient. Here, we define the Semantic Scanpath $\mathcal{S}$ for an interaction turn as the combination of the transcribed utterance of the speaker (dialogue line) and the corresponding gaze history over the temporal window $W$ associated with that turn. The gaze history is represented as an ordered sequence of fixation segments, i.e., pairs containing gazed semantic Areas-Of-Interest (AOIs) and their associated dwell times (considered as a cue of deictic engagement). Formally, ${\mathcal{S} = (u, \mathcal{H})}$, where $u$ denotes the utterance produced during the turn and $\mathcal{H} = \{(A_k, \tau_k)\}_{k=1}^{K}$ denotes the gaze history, with $A_k \subseteq \mathcal{O}$ representing the semantic AOI during segment $k$ (corresponding to one or more objects in the scene) and $\tau_k$ its dwell time. The representation is agnostic as to the way the AOIs sequence is extracted. Accordingly, the multimodal grounding problem can be expressed in terms of this representation as $\mathcal{O}^* = \Phi(\mathcal{S}, \mathcal{O})$. 

In this implementation, AOIs are obtained by associating the gaze direction with the objects in the scene over time. At each sampling instant, candidate objects are ranked according to their angular deviation from this direction, with smaller deviations indicating more likely foci of attention. Fixation segments are then defined as pairs containing the list of most likely fixated objects $A_k$ and the respective dwell time $\tau_k$. Fixations are identified by comparing ranked object lists across consecutive sampling instants. A segment is formed when an object or set of objects remains within a predefined angular threshold $\theta_{th}$ for a minimum duration $T_{th}$, and it ends when the set of fixated objects changes or when a rapid shift in gaze direction occurs, i.e., $A_k=\{o_{\theta_1},\ldots,o_{\theta_n}\}_{\tau_k}$, where $\theta_1\leq\ldots\leq\theta_n<\theta_{th}$ and the associated dwell time satisfies $\tau_k\geq T_{th}$. To handle saccades, object ranking is computed only when the angular velocity of the gaze direction is below a predefined threshold. Consecutive segments referring to the same set of objects and separated by a short temporal interval are merged to avoid fragmentation. 

\begin{figure}[!h]
      \centering
      \includegraphics[width=0.42\textwidth]{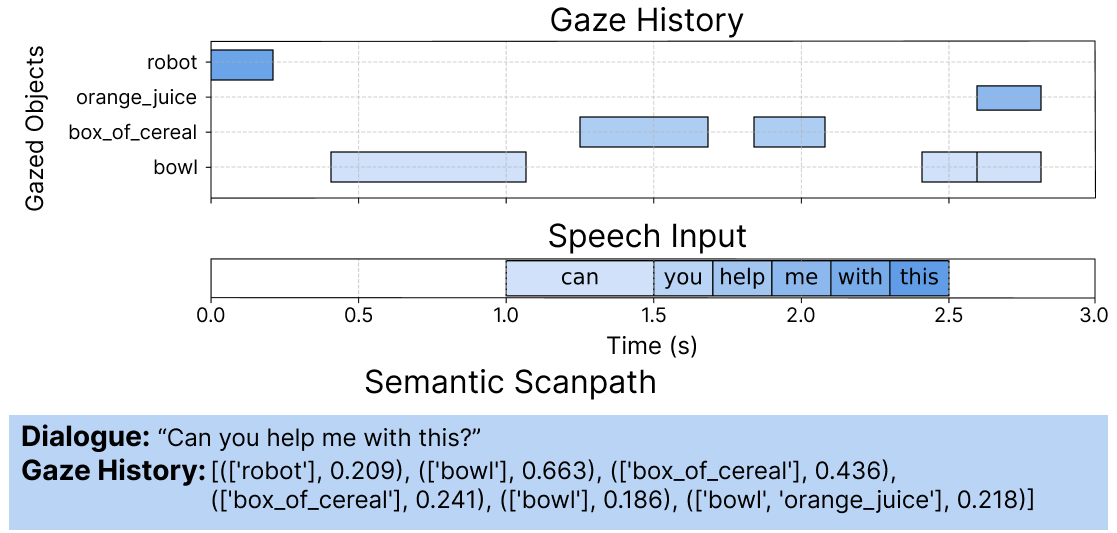}
      \caption{Visualization of gaze history and speech input over time, and the corresponding Semantic Scanpath. The top plot shows the gaze history, capturing fixation segments with durations, while the middle plot presents speech with word-level timestamps. The bottom part presents the Semantic Scanpath, which combines spoken utterances and gaze history.}
      \label{fig:gaze_speech_timeline}
\end{figure}

Fig.~\ref{fig:gaze_speech_timeline} illustrates the gaze history within a turn window (top), the aligned utterance with word-level timestamps (middle), and the resulting textual Semantic Scanpath representation combining speech and gaze (bottom). In our representation, gaze history and the speech utterance are integrated asynchronously, allowing references to be resolved using earlier or later fixations within the turn.


\subsection{Embodied LLM-based agent}
The multimodal grounding mechanism $\Phi$ is realized as an embodied LLM-based agent that processes the Semantic Scanpath $\mathcal{S}$ as input and controls the physical robot.
The robot's character is defined by a system prompt using natural language, prescribing its behavior as a friendly and interactive assistant. 
It is instructed to infer the user’s intent based on speech, gaze history, and context, using gaze history to resolve ambiguities in speech and vice versa. The robot responds through speech and provides explicit reasoning for each decision made. If it is unsure about the user request, it asks for clarification. Finally, the system prompt includes tips for interpreting gaze, emphasizing the frequency and duration of fixations to distinguish referred objects from spurious fixations. For the complete system prompt, see Listing \ref{listing:system_prompt}.

\subsection{Robot's Tools}
The robot's capabilities are controlled by the LLM via an API that exposes external tools, implemented as function calls, and categorized as follows:
\begin{itemize}
    \item \textit{Query Tools} help the robot gather information about its surroundings, including the objects 
    in the scene.
    \item  \textit{Diagnostic Tools} show how the LLM interprets user inputs. It includes the \textit{reasoning} function, which explains the intent inferred using gaze and speech, and the \textit{required\_objects} function, which specifies the objects needed to fulfill the request.
    \item \textit{Expression Tools} enable the robot to communicate its actions and intentions primarily through speech. 
    \item  \textit{Action Tools} enable the LLM to control the robot's manipulation actions in the scene, such as passing an object to the user.
\end{itemize}
See Table \ref{table:tools} for an overview of the available tools, 
their descriptions and corresponding arguments.
Based on the robot's character and capabilities, the agent responds to the Semantic Scanpath input, deciding which tools to invoke. For each interaction turn, the invoked tools and their
outputs were logged for subsequent analysis. These tools are used to inspect the current scene, provide insight into the LLM’s interpretation of the request, communicate with the user, or perform physical actions required to carry out the request. 

\section{EXPERIMENTS AND RESULTS}

\subsection{Experimental methods}
\label{subsec:experimental_methods}
To validate our approach, we conducted an exploratory study to assess how accurately the system works across tasks (requests), scenarios, and users. 

\par{\textbf{Technical setup.}} 
The experimental setup included a microphone, a speaker, and an Azure Kinect camera. The microphone captured user speech, processed by Google Speech-to-Text
. Object poses and identities were estimated using ArUco markers. Additionally, the camera tracked the user's posture, including head orientation. In this work, head orientation was used as a proxy for gaze direction \cite{Andrist2017}. 
At each sampling instant, we computed the head's forward direction and evaluated its angular deviation with respect to each object in the scene using the estimated Axis‐Aligned Bounding Box (AABB). For each object, the minimum angular deviation between the head direction and sampled surface points on the AABB was computed. Objects were then ranked accordingly, with the smallest deviations indicating the most likely foci of attention. Gaze history was then obtained using an angular threshold $\theta_{th}=8^{\circ}$, a fixation duration threshold $T_{th}=100$~ms, a distance between surface points of 5~mm, and a temporal merging window of 160~ms. The same framework could be instantiated with eye-tracking or camera-based gaze target detection methods \cite{tonini2023object,hanifi2024pipeline}.

As the LLM, we used OPENAI's gpt-4-0125-preview, operating with a temperature of $10^{-8}$. It received Semantic Scanpath input for each turn, where gaze history entailed fixation segments detailing viewed objects and fixation durations. The LLM called the tools, and the results from these tool executions were also fed back to the LLM. This evaluation did not involve the action tools, as the focus was on how the LLM resolves ambiguities in user requests by using the proposed representation. While the system only speaks out what it intends to do during data collection, Section \ref{sec:robot} presents a qualitative evaluation where the robot executes the action tools inferred from user requests.

\begin{table*}[t]
\begin{threeparttable}
\setlength{\tabcolsep}{5pt}        
\caption{Overview of the tasks used in the experiments*}
\renewcommand{\arraystretch}{0.9}  

\begin{tabular}{@{}%
  p{1.1cm}  
  p{1.7cm}  
  p{3.0cm}  
  p{2.0cm}  
  p{1.9cm}  
  p{3.0cm}  
  p{2.9cm}  
@{}}
\toprule
\textbf{Scenario/ Task nr.} &
\textbf{Abstract task} &
\textbf{Speech input} &
\textbf{Intended reasoning} &
\textbf{Target objects} &
\textbf{Distractors} &
\textbf{Irrelevant objects} \\
\midrule

Breakfast/\newline T1 &
Infer user \newline intent &
“Can you help me with \underline{this}?” &
Pour cereal in the bowl &
Cereal box, bowl &
Bottle of orange juice, bottle of milk, small bowl &
-- \\
\midrule

Breakfast/ T2 &
Disambiguate object &
“[I’d like to prepare my cereals,] could you pass me \underline{that bottle}?” &
Pass the milk bottle &
Bottle of milk &
Bottle of orange juice &
Cereal box, small bowl, bowl \\
\midrule

Breakfast/ T3 &
Infer content &
“Can I also have some \underline{sugar}?” &
Get the sugar bowl &
Small bowl &
Bottle of milk, bottle of orange juice, cereal box &
Bowl \\
\midrule

Drink/ \newline T1 &
Infer user's \newline preference &
“I’m thirsty, can I have \underline{a drink}?” &
Preference for a cola &
Bottle of cola &
Bottle of cola zero &
Red glass, blue glass, bowl \\
\midrule

Drink/ \newline T2 &
Disambiguate object &
“[The cola, please.] Could you use \underline{this glass}?” &
Use the glass in front of the user &
Red glass &
Blue glass &
Bottle of cola, bottle of cola zero, bowl \\
\midrule

Drink/ \newline T3 &
Infer content &
“I’d like to have some \underline{ice cubes} with it.” &
Get the bowl with ice &
Bowl &
Blue glass &
Bottle of cola zero, bottle of cola, red glass \\
\bottomrule
\end{tabular}

\begin{tablenotes}[flushleft]
\footnotesize
\item[*] In each scenario, the user made three requests to the robot, with different disambiguation goals (as indicated by underlining in the speech input). For each task, the intended reasoning reflects the expected interpretation of the request given the speech input and the user's concurrent scanpath. For each scene and task, target objects were defined as those required to fulfill the request; distractor objects were those that, depending on the context, could be mistaken as the semantic target of the request; and irrelevant objects were not semantically related to the request.
\end{tablenotes}

\label{tab:tasks}
\end{threeparttable}
\end{table*}

\par{\textbf{Procedure.}} For the experimental validation, we devised multiple disambiguation and reference resolution challenges. Besides demonstrative disambiguation (e.g., "this/that" item), we investigated more situated requests, where gaze could reveal an action specification, a hinted preference, or a kind of visuo-verbal metonymy, where the speech indicates the content (not visible to the robot) and the gaze suggests the container for the intended object. Each scenario consisted of a table with 5 objects, with the robot and the human at opposite sides. For the \textit{Breakfast} scenario, a cereal box, a bottle of orange juice, a bottle of milk, a small bowl for the sugar, and a large bowl were placed in front of the user, without occlusions from their perspectives. For the \textit{Drink} scenario, a bottle of Cola and one of Cola Zero were used, along with a red glass, a blue glass, and a bowl (containing ice cubes). Within scenarios, objects were in the same locations across users, not to add further variability. For each scenario, the three sequential requests and the corresponding intended robot reasoning
, along with the categorization of involved objects, are presented in Table \ref{tab:tasks}. 
The data collection involved 7 people (3 female, 4 male) recruited among coworkers, each interacting with the system once across the two different scenarios. We aimed 
at collecting data from realistic interactions and with enough variability to validate the system robustness. Users were instructed to look at the camera as if it were the robot and to direct their gaze toward the relevant objects while uttering the given request. No specific constraints were imposed on their gaze behavior
\footnote{Before starting, users were familiarized with the gaze detection modality by asking them to look at randomly chosen objects as if pointing at them with their nose. The system gave then feedback when the object was hit.}. Each interaction was defined by the time window of the user's turn: it began when participants established gaze with the robot before speaking (signaled by the experimenter with a key press that also triggered an auditory cue) and ended when the utterance was completed (marked by another key press). Participants were instructed to look back at the robot after finishing, but this did not always occur; in all cases, the window was consistently closed at utterance completion by a second key press. This practical choice ensured consistency across users; in future implementations, such time windows could be detected automatically by combining speech activity detection with turn-taking gaze behavior. Since the system was reacting in real-time (although not physically acting) and the LLM could ask for clarification in case of uncertainty (e.g., "Would you like the cola or the cola zero?"), users might confirm the previous target before making the following request, so to maintain the interaction coherent ("The cola please, and can you use this glass?").  During each turn, we recorded raw gaze data, gaze history, and speech inputs and stored the system responses. If the target objects were not properly hit with the gaze and would not appear in the gaze history, the user was asked to repeat the request to ensure valid multimodal input. 
In this way, we collected 6 turns for each user (2 full interactions with 3 requests). To expand this dataset for offline evaluation, we applied a combinatorial approach, where the inputs from different users were mixed to generate new synthetic interaction dialogues. With 3 tasks and 7 users, this yielded 343 possible T1-T3 interaction combinations for each scenario. This accounts
for the LLM's stochasticity in replying to the very same input and for sequence effects in the interactions.  
To investigate the role of the different input modalities and of specific Semantic Scanpath features, we ran different control versions of the system on the same dataset:
\begin{enumerate}
    \item \textit{Speech + scene:} only the utterance and the tool to query the scene were fed to the LLM. This baseline shows
    the value of adding gaze information to the requests.
    \item \textit{Shuffled gaze + speech + scene}: in the gaze history, the fixation sequence is randomly shuffled.
    \item \textit{Gaze (no dwell times) + speech + scene}: in the gaze history the fixation duration is removed.
    \item \textit{Gaze (only first object) + speech + scene}: here for each fixation in the gaze history only the most likely object is provided, along with the duration.
    \item \textit{Semantic Scanpath}: the full Semantic Scanpath without tools to query the scene.
    \item \textit{Semantic Scanpath + scene}: full Semantic Scanpath with scene querying.
\end{enumerate}

Each version had an adapted system prompt, where the rules or tips concerning gaze interpretation or fixation duration were removed, if these were not part of the input. For the Semantic Scanpath we evaluated a version without scene querying to assess whether the LLM could use such information to rule out that the user was referring to objects present in the scene but not glanced at, in case of uncertainty about the request.
The responses from the LLM's diagnostic tools were collected and analyzed to assess the system's performance in resolving ambiguous requests.

\begin{figure}
    \centering
    \includegraphics[width=0.94\linewidth]{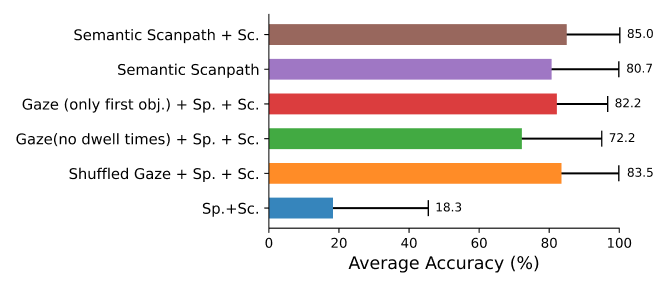}
    \caption{Accuracy for each condition with respect to the ground truth inference averaged across tasks and scenarios. Black lines represent standard deviation.}
    \label{fig:accuracy}
\end{figure}


\begin{table}[ht]
\centering
\caption{Accuracy (\%) of the different conditions across tasks. Abbreviations: Gz. = gaze; Sp. = speech; Sc. = Scene}
\renewcommand{\arraystretch}{1.3} 
\begin{tabular}{p{3.3cm}| p{0.36cm} p{0.36cm} p{0.36cm} |p{0.36cm} p{0.36cm} p{0.36cm}}
\toprule
 & \multicolumn{3}{c|}{\textbf{Breakfast}} & \multicolumn{3}{c}{\textbf{Drink}}\\
\textbf{Condition} & T1 & T2 & T3 & T1 & T2 & T3 \\
\midrule
1) Sp. + Sc.& 0.0& 77.6 & 0.0 & 8.8& 18.1 & 5.2 \\
2) Shuff. gaze + Sp. + Sc. &90.4 & 98.5 & 50.7 & 85.4 & 99.7& 76.4\\
3) Gz. (no dwell) + Sp. + Sc. & 80.8 & 99.1 & 35.6 & 71.4 & 95.0 & 51.0 \\
4) Gz. (first obj.) + Sp. + Sc. & 75.5 & 99.8 & \textbf{58.9} & \textbf{85.7} & 99.4 & 74.6 \\
5) Semantic Scanpath & 90.4 & 99.4 & 45.5 & 84.2 & 98.8 & 65.8 \\
6) Semantic Scanpath + Sc. & \textbf{91.6} & \textbf{100.0} & 56.3 & \textbf{85.7} & \textbf{100.0} & \textbf{76.7} \\
\bottomrule
\end{tabular}
\label{tab:accuracy}
\end{table}
\begin{figure*}
    \centering
    \includegraphics[width=0.72\textwidth]{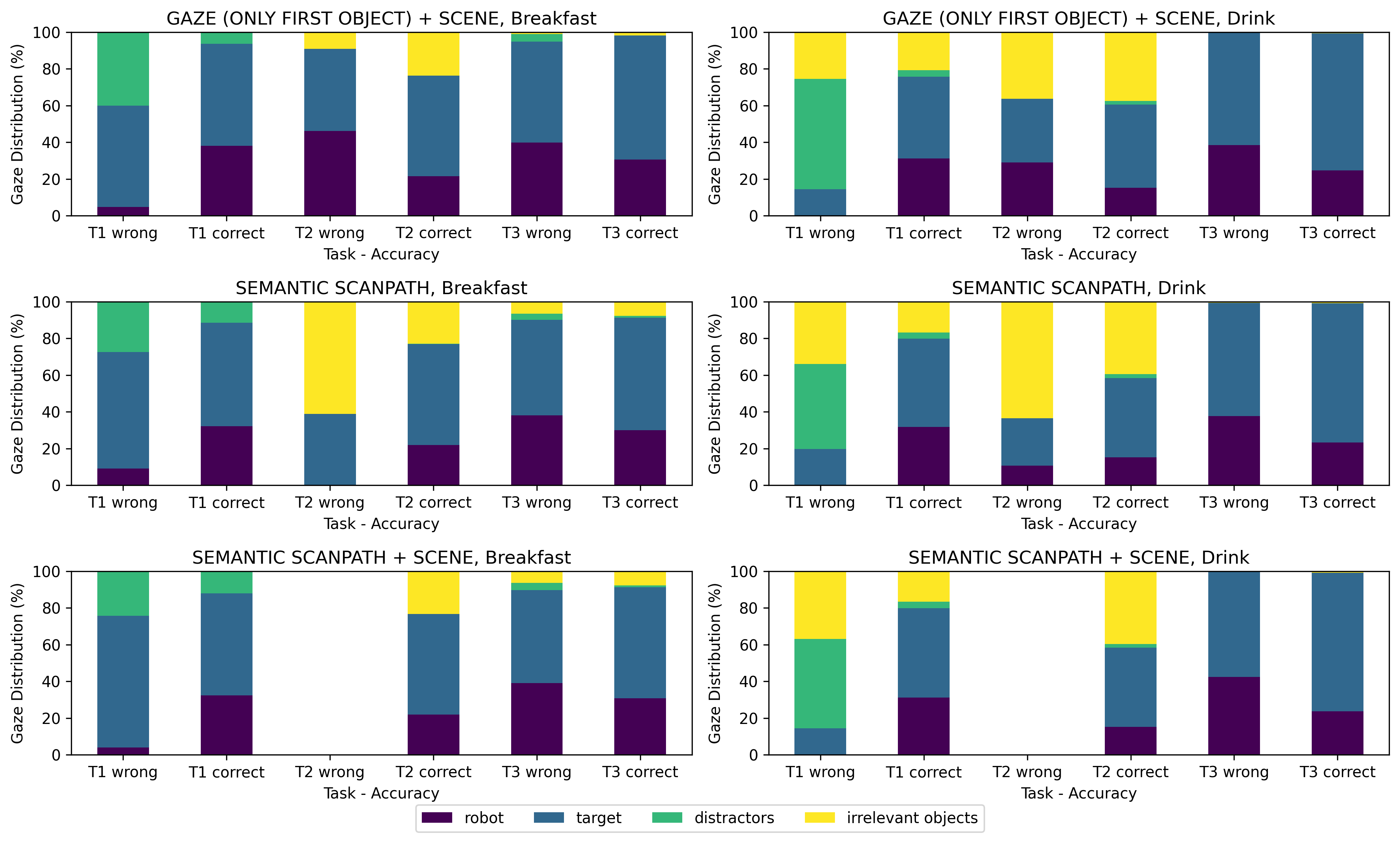}
    \caption{Gaze distribution per task (T1-T3) and scenario (breakfast, drink) across relative categories of objects for all conditions with gaze and dwell times. In cases where the system accurately inferred the user intent, the gaze dwelled primarily on the target objects, with some misleading fixations on irrelevant or distractor objects. Note that T2 for the Semantic Scanpath and scene condition was always correctly resolved, thus T2 wrong distribution appears empty.}
    \label{fig:gazedistro}
\end{figure*}

\subsection{Results}

We first looked at the accuracy across tasks in the two scenarios. The system reply was deemed correct when the \textit{'required\_objects'} tool output contained the target object(s) as presented in Table \ref{tab:tasks}. This was deemed a more objective behavioral metric, considering that reasoning output could be plausible but not faithful \cite{madsen2024self}. Accuracy was computed at the object level, as the study focused on reference resolution and the prompt specifically instructed the LLM to ask back, in case it was unsure about the user's intent. Overall accuracy results are presented in Fig. \ref{fig:accuracy} and in detail in Table \ref{tab:accuracy}. The condition combining the Semantic Scanpath and the possibility to query the scene achieved the best accuracy, scoring significantly higher than every other condition (McNemar's test with Holm-Bonferroni correction for multiple comparisons, all $\chi^2(1)>7.04$, all $p < .01$). In disambiguating demonstratives, this condition reached ceiling performance, and performed best in most tasks (see bold values in Table \ref{tab:accuracy}). The shuffled gaze condition achieved the second-best performance, suggesting that, in this dataset, the correctly ordered sequence of fixations had a limited impact relative to availability of gaze and scene information. Scene querying 
grounds the system interpretation in the detected objects, preventing hallucinations of non-existent candidates (as in both T3 tasks). Overall, the superior performance of the Semantic Scanpath + scene condition w.r.t. other conditions suggests that dwell time,  the possibility of multiple candidate objects within a fixation, and scene querying provided valuable cues for disambiguating user requests, while fixation order had a more limited effect in these tasks.

Another variable that can influence the system's accuracy is the degree of ambiguity in the gaze history itself.
While users were asked to look at the intended objects, they were not required to avoid glancing at other objects. This would have been an unnatural constraint since objects were spread across the table and users' line of sight was not constantly aligned with their head direction. The ultimate goal was to have a relatively normal speech and gaze interaction with the robot. To assess the role of gaze behavior, we computed the mean percentage distribution of gaze time across trials for the different categories of visual objects in the scene (the camera/robot, the task targets, the distractors, and the irrelevant objects). For gaze segments where two different objects could be the target, the dwell time was evenly split among the corresponding categories. Across participants, only 20\% of fixation segments contained more than one object. We then compared the distributions for wrongly and correctly inferred tasks, in the versions with gaze and dwell time (see Fig. \ref{fig:gazedistro}). Across conditions, when the system was wrong, the gaze lingered more on distractors and irrelevant objects, as expected. For T3 in both scenarios, the proportion of time spent on the target appears to have helped resolve the container ambiguity. Still, in some cases, the system was correct even when the gaze partially lingered on distractors and irrelevant objects.

\begin{figure*}[!t]
      \centering
      \includegraphics[width=0.72\textwidth]{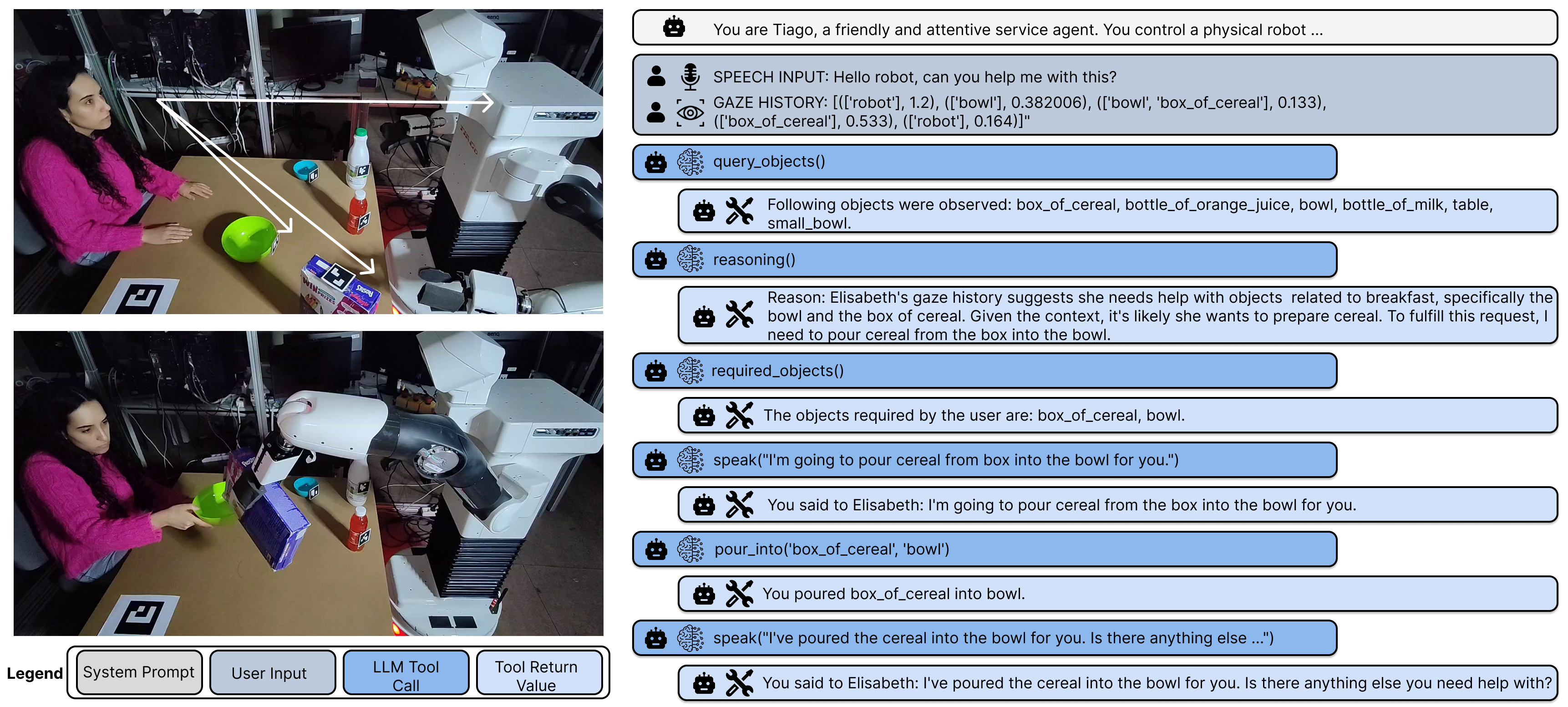}
      \caption{Proof-of-concept robot demonstration showing how the LLM integrates gaze history and speech input to infer the intended task ("Can you help me with this?", resolved as pouring cereal into the bowl). Left: Snapshots of key moments of the interaction and the robot’s execution. Right: Sequence of tool calls made by the LLM, including query tools, diagnostic tools (e.g., \textit{reasoning}), expression tools (e.g., \textit{speak}), and the final action tool \textit{{pour\_into}}.}
      
      \label{fig:robot_demonstration}
\end{figure*}

\subsection{Robot Demonstration}\label{sec:robot}

We conducted a robot demonstration to qualitatively illustrate how the interpreted 
user intent can guide physical actions. This demonstration was designed as 
proof-of-concept, showing the feasibility of integrating our approach on a physical platform with real action capabilities rather than serving as a systematic evaluation\footnote{The full implementation used in this work, including code, prompts, and tools is publicly available at \url{https://github.com/elisabeth-ms/SemanticScanpath.git}}.
The demonstration uses the TIAGo++ robot
, controlled via ROS Noetic, with action tools implemented as a ROS action client. These high-level abstractions, similarly to \cite{Tanneberg2024}, translate the LLM’s interpreted commands into robot actions. An external Intel RealSense camera 
captured both the scene and the user, addressing the limited field of view of TIAGo++’s head camera.
The gaze history was computed 
as in the user experiments, and the spoken utterance was transcribed with Google Speech-to-Text.  Fig. \ref{fig:robot_demonstration} presents an example in which the user asks "Can you help me with this?" while looking at the bowl and the cereal box. The LLM interprets the request, chooses the necessary tools, and commands the robot to pour cereal into the bowl using the \textit{pour\_into} action tool.

\section{DISCUSSION}
Our results support the generality and flexibility of the Semantic Scanpath representation in prompting an LLM. The gaze history is provided in text format together with the utterance, which makes it more efficient than scanpath visualization on an image or sequence of images. The provided system prompt was fine-tuned as to the robot's character, description of input format, perceptual and action capabilities, and complemented with general suggestions on interpreting the scanpath, but not tailored to a specific scenario or task. In our case, thus, the gaze was not used just as a pointer but as an additional input with its temporal dynamics, loosely aligned with the speech. By sampling the output of the \textit{reasoning} tool in the Semantic Scanpath conditions, we observe that LLMs appear able to reason about Semantic Scanpaths, not just to disambiguate demonstratives by referential gaze, but to discriminate the truly intended objects by considering which objects were looked at and for how long. For example, in T1 in the drink scenario, even when both bottles appeared in the gaze history, the model could reason that \textit{"The user's gaze history indicates a longer and more focused attention on the bottle of cola compared to the bottle of cola zero, suggesting a preference for the bottle of cola when she mentioned wanting a drink."}. While again, LLMs' reasoning explanations should be interpreted cautiously \cite{madsen2024self}, such motivations suggest, on the one hand, that the Semantic Scanpath representation allows LLMs to interpret gaze-based intentions, or even preferences, without the need to formalize specific tasks and train machine learning models for scanpath analysis \cite{mohamed2024review}. On the other hand, they show the potential to reason about both speech and gaze semantics concurrently and weigh those cues for disambiguation also in relation to the context and scene (e.g., \textit{"The user's gaze history shows a focus on a small bowl, which likely contains sugar, given the context of preparing cereal. Her request for sugar and the gaze on the small bowl suggest she needs the small bowl passed to her."}) While we focused on inferring the requested objects, in T1 in the breakfast scenario, the request entailed also an ambiguous action. The prompt prescribed the LLM to follow up with a question in case of uncertainty. We thus considered how often a question was asked in the robot's reply: this was virtually always the case for this task (99.95\%), yet the inference was wrong only in 28.6\% of the cases. Often the robot identified the correct objects but replied asking something like \textit{"Do you need help with preparing cereal?"}. In general, the worst ratio of replies with questions, after a wrong inference, to overall replies with questions occurred in both T3 tasks (over 60\%), suggesting that those were the tasks where the robot was most uncertain about the inference, and questions were used to resolve that.
The Semantic Scanpath-based approach thus offers a promising perspective to achieve shared understanding and common ground in situated human-robot interaction, as previously suggested \cite{Chai2014,gross2017reliability}.
The study provides evidence for the efficacy of the Semantic Scanpath representation in resolving ambiguous and underspecified references, although some limitations remain in the proposed implementation. The obtained results are influenced by the accuracy in gaze estimation and by relying on fiducial markers to detect a finite number of labeled objects. Accuracy could be improved by more reliable state-of-the-art gaze tracking methods. The closed-world assumption could be further lifted by using the latest segmentation and VLM models to retrieve present objects and their labels. The same representation could in principle be used in that setting. 

Future work should investigate the effect of such disambiguation capabilities on naive users with a user study. The generalization capabilities to different task domains and pragmatic implicatures \cite{sravanthi-etal-2024-pub} also need to be more systematically explored. As for understanding indirect speech acts \cite{zhang2025can}, we expect that understanding gaze and speech references would enhance trust and goal alignment in human-robot collaboration, but the tasks and contexts where this is the case need to be experimentally assessed.

\section{CONCLUSIONS}
In this work we proposed a compact and semantically grounded representation combining the user's verbal and gaze behavior to enable LLM-powered robots to resolve ambiguous or underspecified references. We provided a tailored but general prompt to appropriately interpret such representation and demonstrated the accuracy and flexibility of the approach across different tasks, scenarios, and user gaze behavior. A qualitative robot demonstration further showed a straightforward integration with an embodied agent able to execute the inferred requests. 

\section*{ACKNOWLEDGMENT}

This research has been supported by project  iRoboCity2030-CM, Robótica inteligente para ciudades sostenibles (TEC-2024/TEC-62), funded by Programas de Actividades I+D en tecnologías de la Comunidad de Madrid. AI tools were used for language editing; all content was reviewed by the authors.


\begin{table*}[b]
\centering
\caption{Overview of Available Tools and Their Arguments}

\label{table:tools}

\begin{tabular}{l p{0.31\linewidth} p{0.43\linewidth}}
\hline
\textbf{Tool} & \textbf{Description} & \textbf{Arguments} \\ 
\hline
\multicolumn{3}{l}{\textbf{Query Tools}} \\ 
\hline
query\_objects & Query all objects that are available in the scene. You can see all these objects. & - \\ 
\hline
\multicolumn{3}{l}{\textbf{Diagnostic Tools}} \\ 
\hline
reasoning &  You provide a reason for the action you are about to take. & - \\ 
required\_objects &  You provide the name of the objects the user requires to fulfill a request. & - \\ 
\hline
\multicolumn{3}{l}{\textbf{Expression Tools}} \\ 
\hline
speak &     You speak out the given text. & \textbf{person\_name}: The name of the person to speak to. \\  
      &  & \textbf{text}: The text to speak. \\  
\hline
\multicolumn{3}{l}{\textbf{Action Tools}} \\ 
\hline
move\_object\_to\_person & You get an object and move it to a person. & \textbf{object\_name}: The name of the object to move. The object must be available in the scene. \\  
                          &  & \textbf{person\_name}: The name of the person to move the object to. \\  
\hline
hand\_object\_over\_to\_person &You get an object and hand it over to a person. & \textbf{object\_name}:The name of the object to hand over. The object must be available in the scene. \\  
                                &  & \textbf{person\_name}: The name of the person to hand over the object to.\\  
\hline
pour\_into & You get a source container, pour it into a target container, and put it back on the table. & \textbf{source\_container\_name}: : The name of the container to pour from. \textbf{target\_container\_name}:  The name of the container to pour into. \\  
\hline
\end{tabular}
\label{tab:tools}
\end{table*}

\IEEEtriggeratref{36}
\bibliographystyle{IEEEtran_etal}
\bibliography{gaze_disambiguation}
\section*{APPENDIX}
See Listing \ref{listing:system_prompt} for the system prompt and Table \ref{tab:tools} for the available tools.

\begin{figure}[h]
\begin{lstlisting}[caption={System prompt.},captionpos=t, label={listing:system_prompt}, basicstyle=\scriptsize\ttfamily]
You are a friendly and attentive service agent.
You control a physical robot called 'the_robot' and receive requests from the user.
You have access to functions for gathering information, acting physically, and speaking out loud.
You receive two types of inputs from the user:
    Dialogue: The user will verbally ask for help.
    Gaze history: This is divided into segments, each showing the objects the user likely focused on while uttering the speech input and the duration of that focused period (seconds). Some segments may include multiple objects ordered by decreasing likelihood (closer objects are mixed). 
IMPORTANT: Obey the following rules:
1. Always start gathering all available information related to the request from the scene and the input. 
2. Always focus on understanding the user's intent based on context, speech input, and gaze history. Use gaze to clarify speech, when requests are ambiguous. Use speech to clarify gaze, when requests are ambiguous.
3. Provide a reason for every response to user requests using the 'reasoning' function to explain decisions. Be concise and clear.
4. Speak out loud using the 'speak' function to communicate clearly and concisely with the user.
5. If you are not sure about the user's intent, ask for clarification.
6. Provide the 'required_objects' for every user request.
REMEMBER YOUR RULES!! TIPS FOR INTERPRETING GAZE:
1. Referred objects are usually gazed ahead of utterance, but also right before looking at you.
2. Intentionally referred objects are usually looked at longer and more frequently.
3. Spurious fixations are usually short and mixed with closer objects.
\end{lstlisting}
\end{figure}

\end{document}